\documentclass[twocolumn,prb,aps,showpacs]{revtex4}
\newcommand{\BEQ}{\begin{equation}}     
\newcommand{\BEA}{\begin{eqnarray}}
\newcommand{\EEQ}{\end{equation}}       
\newcommand{\EEA}{\end{eqnarray}}
\newcommand{\D}{{\rm d}}

\begin{document}

\input epsf.sty



\title{Aging phenomena in critical semi-infinite systems}

\author{Michel Pleimling}

\affiliation{
Institut f\"ur Theoretische Physik I, Universit\"at Erlangen-N\"urnberg,
D -- 91058 Erlangen, Germany}

\begin{abstract}
Nonequilibrium surface autocorrelation and autoresponse functions are studied numerically
in semi-infinite critical systems in the dynamical scaling regime. Dynamical
critical behaviour is examined for a nonconserved order parameter
in semi-infinite two- and three-dimensional Ising models as well as in the
Hilhorst-van Leeuwen model. The latter model permits a systematic study of
surface aging phenomena, as the surface critical exponents change continuously
as function of a model parameter. The scaling behaviour of surface two-time quantities
is investigated and scaling functions 
are confronted with predictions coming from the theory of local
scale invariance. Furthermore, surface fluctuation-dissipation ratios are computed
and their asymptotic values are shown to depend on the values of surface
critical exponents. 
\end{abstract}
\maketitle

\section{Introduction}
Aging behaviour observed in systems with slow degrees of freedom
is one of the most intriguing aspects in nonequilibrium physics,
see Refs.\ \onlinecite{God02,Cug02,Cri03}
for recent reviews.
This behaviour is due to relaxation processes which depend on the 
thermal history of the studied sample. Aging phenomena have been 
studied extensively in disordered systems like spin glasses or
glassy systems, but they are also encountered in the
simpler ferromagnetic systems.\cite{Bra94,God02}
In the last years remarkable progress
has been achieved in our understanding of the physics of aging phenomena
taking place in ferromagnets.\cite{Cug94,Bar98,Ber99,God00a,God00,Lip00,Hen01,Ber01,Cal02,
Hen02,Pic02,May03,Sas03,Hen03c,Hen03,Hen03a,Maz04,Abr04,God04,Pic04,Cha04,Abr04b}

In the typical scenario the ferromagnetic system is prepared in a fully disordered state
at high temperatures and then quenched down to temperatures $T \leq T_c$,
where $T_c > 0$ is the critical temperature. When the final temperature is
smaller than $T_c$, coarsening takes place, leading to the formation of domains
with a time-dependent typical length. It is the slow
motion of the domain boundaries which is responsible for the aging processes. 
The situation is different for a quench to the critical point, where one
is dealing with nonequilibrium critical dynamics, as ordered
domains do not exist in that case. The typical
time-dependent length is then given by the dynamical correlation length
$\xi(t) \sim t^{1/z}$, where $z$ is called the dynamical critical exponent.

In recent years investigations concentrated on two-time quantities like
dynamical correlation and response functions, as they more fully reveal
the aging processes. Examples are the spin-spin autocorrelation function $C(t,s)$
and the conjugate response function $R(t,s)$ defined by
\BEQ
C(t,s) = \left\langle \phi(t) \phi(s)\right\rangle \;\; , \;\;
R(t,s) = \left.\frac{\delta \langle\phi(t)\rangle}{\delta h(s)}\right|_{h=0}.
\EEQ
Here $\phi(t)$ is the time-dependent order parameter, whereas $h(s)$
is the magnetic field conjugate to $\phi$. The time $t$ elapsed since the quench
is called the observation time and $s$ is referred to as the waiting time.
Of course, $R(t,s)=0$ for $t<s$, due to causality.
In the dynamical scaling regime, $t\gg t_{\rm micro}$,
$s\gg t_{\rm micro}$ and $\tau =t-s\gg t_{\rm micro}$, where $t_{\rm micro}$
is some microscopic time, these quantities are expected to have the
following behaviour \cite{God02,Cug02}:
\BEQ \label{1:gl:CRskal}
C(t,s) = s^{-b} f_{C}(t/s) \;\; , \;\;
R(t,s) = s^{-1-a} f_{R}(t/s).
\EEQ
In the limit $y= t/s \to\infty$ the scaling functions $f_{C,R}(y)$
present a simple power-law behaviour
\BEQ
f_{C}(y) \sim y^{-\lambda_C/z} \;\; , \;\;
f_{R}(y)\sim y^{-\lambda_R/z}
\EEQ
with the autocorrelation \cite{Hus89} and
autoresponse \cite{Pic02} exponents $\lambda_C$ and $\lambda_R$.
In the case of a quench to the critical point of a ferromagnetic system 
with initial short-range correlations we have
$a=b=2 x/z$ and $\lambda_C=\lambda_R=d-x_i+x$ where $d$ is the number
of space dimensions, whereas $x$ resp. $x_i$ denotes the scaling dimension
of the order parameter resp.\ of the initial magnetization.\cite{Jan89}
A generalization of dynamical scaling to local space-time dependent scaling
yields the prediction \cite{Hen01,Hen02}
\BEQ \label{eq_lsi}
f_R(y)= r_0 \, y^{1+a-\lambda_R/z} \, (y-1)^{-1-a}
\EEQ
for the scaling function of the response function $R(t,s)$. This prediction
was found to be in excellent agreement with Monte Carlo simulations of
two- and three-dimensional Ising models both at and below
$T_c$.\cite{Hen01,Hen03,Hen03a} Similar agreement has recently also been observed
for the three-dimensional $XY$ model.\cite{Abr04b} On the other hand recent renormalization
group calculations \cite{Cal02} yielded at critical points a correction at two loops to the
prediction (\ref{eq_lsi}) coming from the theory of local scale
invariance.\cite{Hen02} The origin of this discrepancy is still not
elucidated, but it must be noted that the simulations are based on a master equation,
whereas the field theoretical calculations start from a Langevin equation.

Recently, the fluctuation-dissipation ratio
\begin{equation} \label{eq_x}
X(t,s)=T \, R(t,s)/\frac{\partial\, C(t,s)}{\partial \, s}
\end{equation}
has attracted much attention in nonequilibrium systems.\cite{Cri03} 
At equilibrium, both
$C(t,s)$ and $R(t,s)$ depend only on the time difference $t-s$, and $X(t,s)=1$,
thus recovering the well-known fluctuation-dissipation theorem.
However, this is not the case far from equilibrium where $X(t,s) \neq 1$.
This property led to numerous studies of $X(t,s)$ in various systems with
the aim to possibly quantify the distance to equilibrium.
In the context of glassy systems the limit value $X^\infty$ has been
interpreted as giving rise to an effective temperature $T_{eff} = T/X^\infty$.\cite{Cug97}
Godr\`eche and Luck conjectured that in critical systems the asymptotic value 
$X^\infty =
\lim\limits_{s \longrightarrow \infty} \left(
\lim\limits_{t \longrightarrow \infty} X(t,s) \right)$ of this
nonequilibrium quantity
characterizes the different universality classes.\cite{God00} 
This has indeed been confirmed in
exactly solvable cases,\cite{God00a,Lip00,Hen03b} in
various Monte Carlo studies \cite{God02,God00,Cha03,May03,Sas03,Cha04} as well as in
field-theoretical calculations.\cite{Cal02}

Surfaces are unavoidable in real materials and play an outstanding role
in nanotechnologie. It is therefore of importance to understand the local
behaviour close to a surface in nonequilibrium systems.
In this work we take a step in that direction and
extend the investigation of aging phenomena in critical
systems to the semi-infinite geometry, focusing on aging processes
taking place in the vicinity of the surface.

It is well known that the presence
of symmetry-breaking surfaces changes local quantities, leading to surface
critical behaviour which differs from that of the bulk (see Refs.\
\onlinecite{Die97,Ple04a}
for recent reviews). For example, the static critical surface pair correlation  
function behaves as $\left| \vec{\rho}-\vec{\rho}\,' \right|^{-2 x_1}$,
where for a $d$-dimensional model $\vec{\rho}$ is a $(d-1)$-dimensional vector 
parallel to the surface. The scaling dimension $x_1$ of the surface order parameter
differs in general from the scaling dimension $x$ of the bulk order parameter.
Of further importance is the fact that,
depending on the surface interactions,
different surface universality classes may be realized
for a given bulk universality class. An important case is
that of the three-dimensional semi-infinite Ising model where 
three different surface universality classes are encountered 
at the bulk critical temperature.

The present work studies the effect of surfaces on aging processes 
with a non-conserved order parameter in
the dynamical scaling regime $t,s \gg t_{micro}$, $t-s \gg t_{micro}$.
We thereby
discuss not only two- and three-dimensional semi-infinite Ising models, 
but also analyze aging phenomena
taking place in the Hilhorst-van Leeuwen model.\cite{Hil81,Igl93}
The latter model is a semi-infinite two-dimensional Ising model with an extended
surface defect.
In this model the surface scaling dimension $x_1$ has been shown to
vary continuously as a function of the defect amplitude
in the interesting case that the surface
defect is a marginal perturbation. This makes it possible to
systematically study at criticality the impact of surfaces on local two-time 
correlation and response functions
as well as on related quantities as local fluctuation-dissipation ratios.

This work is also meant to close a gap in the study of critical dynamics
at surfaces. Prior works either treated equilibrium critical dynamics near surfaces
\cite{Die83,Die94,Die02} or focused mainly on the effect a surface has on
nonequilibrium critical dynamics immediately after the quench to the critical point.
\cite{Rit95,Maj96,Kik85,Rie85,Ple03} To our knowledge the dynamical scaling regime
has not yet been investigated in semi-infinite critical systems.

The paper is organized in the following way. In Section II the different
semi-infinite models are introduced and some important facts on equilibrium and dynamic 
surface critical behaviour are reviewed. Section III deals with surface two-time
autocorrelation functions in the dynamical scaling (i.e.\ aging) regime, whereas
Section IV is devoted to the corresponding study of the thermoremanent 
magnetization. These quantities are examined in the Hilhorst-van Leeuwen
model for various values of the defect amplitude. In the three-dimensional
semi-infinite Ising model, both the ordinary transition (where the
bulk alone is critical) and the special
transition point (where bulk and surface are both critical)
are considered. In Section IV numerically determined scaling functions 
derived from the
intergrated responses are compared with predicted scaling functions
coming from the theory of local scale invariance. Section V is devoted
to the analysis of surface fluctuation-dissipation ratios. It is shown
that the limit value of this ratio depends on the value of the surface
scaling dimension $x_1$. Interpreting this asymptotic value as giving rise 
to an effective
temperature, it follows that for a given bulk universality class different
surface effective temperatures may be obtained by varying the value of
some local model parameters. Finally, Section VI gives our conclusions
as well as an outlook on open problems.

\section{Critical behaviour near surfaces}
\subsection{Models and surface phase diagrams}
All the models studied in this work are defined on $d$-dimensional
semi-infinite (hyper)cubic lattices. Every lattice point is characterized
by an Ising spin which takes on the values $\pm 1$. For the pure, perfect
Ising model in absence of external fields, the Hamiltonian is given by
\begin{equation} \label{eq_ham1}
{\mathcal H}=- J_b \sum\limits_{bulk} \, \sigma_i \sigma_j 
- \, J_s \sum\limits_{surface} \, \sigma_i \sigma_j .
\end{equation}
The first sum is over nearest neighbour pairs where at most one spin
is a surface spin, whereas the second sum is over nearest neighbour pairs with
both spins lying in the surface layer. $J_b$ resp.\ $J_s$ is the strength of the bulk
resp.\ surface couplings. All couplings are supposed to be ferromagnetic,
i.e.\ $J_b$,  $J_s > 0$.

The surface phase diagrams of the two- and three-dimensional semi-infinite
Ising models (\ref{eq_ham1}) are well established.\cite{Die97,Ple04a} 
In two dimensions only one surface universality
class, the so-called ordinary transition, is encountered at the bulk 
critical point for all values of the surface couplings. This transition
is characterized by the value $x_1 = 1/2$ of the scaling dimension of
the surface order parameter. Here and in the following I follow the usual convention
and characterize surface quantities by the index 1.
The surface phase diagram is more interesting
for the three-dimensional model. If the ratio of the surface coupling $J_s$ 
to the bulk coupling $J_b$, $r=J_s/J_b$,
is sufficiently small, the system undergoes at the bulk critical temperature
$T_{c}$ an ordinary transition, with the bulk and surface orderings occuring
at the same temperature.
Beyond a critical ratio, $r > r_{sp} \approx 1.50$ for the semi-infinite Ising model
on the simple cubic lattice,\cite{Bin84,Rug92}
the surface orders at the so-called surface transition at a
temperature $T_s > T_{c}$, followed by the extraordinary transition
at $T_{c}$. At the critical ratio $r_{sp}$, one encounters the multicritical
special transition point,
with critical surface properties deviating from those at the ordinary transition
and those at the surface transition.
The present work exclusively treats aging processes taking place at the
ordinary transition and at the special transition point. For these cases
the scaling dimension $x_1$ takes on the values $1.26$ (ordinary transition)
and $0.376$ (special transition point).

The Hilhorst-van Leeuwen model \cite{Hil81,Igl93}
is a critical two-dimensional semi-infinite Ising model
with an extended surface defect due to inhomogeneous couplings. Considering
a square lattice, one has 
couplings with a constant strength $J_1$
in the direction parallel to the surface,
whereas the strength of the
couplings varies perpendicular to the surface as a function of the distance $l$
to the surface:
\begin{equation} \label{eq_hvl}
J_2(l) - J_2(\infty) = \frac{\tilde{A}}{l^\omega}.
\end{equation}
with $\tilde{A}=A \, T_c \sinh (2 J_2(\infty)/T_c)/4$.\cite{Blo83} 
In this work $J_1$ is set equal to $J_2(\infty)$.
This extended perturbation is irrelevant for $\omega > 1$, yielding
the same critical behaviour as the homogeneous semi-infinite system. For $\omega < 1$ and $A>0$,
however, the perturbation is relevant and a spontaneous surface magnetisation
is observed at the bulk critical point. In the present context 
the most interesting case is obtained for $\omega =1$,
where the perturbation is marginal for $A<1$ (for $A>1$ one again observes
a spontaneous surface magnetisation). Indeed, exact results 
show that the scaling dimension $x_1$ is then a continuous function of the
parameter $A$ with
\begin{equation} \label{eq_hvl2}
x_1 = \frac{1}{2} ( 1- A)
\end{equation}
and $A < 1$.
This intriguing behaviour, which has attracted much interest in the past,
gives us the possibility to continuously change the surface critical exponents
for a given bulk universality class (that of the two-dimensional Ising model),
thereby making a systematic study of aging processes at critical surfaces possible.

\subsection{Critical dynamics near surfaces}
Besides changing the local static critical behaviour, surfaces have also an effect on
critical dynamics. This has been studied in semi-infinite extensions
of the well-known bulk stochastic models,\cite{Hoh77} 
as for example model A (purely relaxational
without any conserved quantities) \cite{Die83,Die94} 
or model B (with conserved order parameter).\cite{Die92,Die94}
A central aspect of the works on dynamic surface critical behaviour
concerns the possible classification of the distinct surface dynamic universality
classes.

\begin{table}
\caption{Literature values of bulk and surface scaling dimensions $x$ and $x_1$
 as well as of the nonequilibrium
dynamical bulk exponents $x_i$ and $z$, for both
the two- and the three-dimensional Ising models.
OT: ordinary transition, SP: special transition point. \label{table_1}}
\begin{tabular}{|c|c|c|c|c|}  \hline
 & $x$ & $x_1$ & $x_i$ & $z$ \\ \hline
$d=2$ & $1/8$ & $1/2$ & $0.53$ & $2.17$ \\
$d=3$, OT & $0.516$ & $1.26$ & $0.74$ & $2.04$ \\
$d=3$, SP & $0.516$ & $0.376$ & $0.74$ & $2.04$ \\ \hline
\end{tabular}
\end{table}   	  

It is important to note that no genuine dynamic surface exponent exists.\cite{Die83,Rit95}
All exponents describing the surface critical behaviour of dynamic quantities can be
expressed entirely in terms of static bulk and surface exponents and the dynamic
bulk exponents $z$ and $x_i$. The values of the critical exponents that are
of interest in the following are listed in Table~\ref{table_1}  for the two- and the
three-dimensional Ising models.
Consider as an example the
dynamic surface spin-spin autocorrelation function. At equilibrium it decays for
long times as $t^{-2 x_1/ z}$,\cite{Die83} whereas out-of-equilibrium one expects
from scaling arguments and field-theoretical calculations the
power-law behaviour $t^{-\lambda_1/z}$ with the surface autocorrelation exponent
\begin{equation} \label{eq_lambda1}
\lambda_1=\lambda_C+2(x_1-x)
\end{equation}
and $t \gg 1$ being the time elapsed since the 
quench.\cite{Rit95,Maj96} This prediction has up to now only been verified
in semi-infinite Ising models.

Out-of-equilibrium studies of dynamic surface critical behaviour are very scarce.\cite{Rit95,Maj96,
Kik85,Rie85,Ple03}
In Ref.\ \onlinecite{Rit95} short-time critical relaxation was analysed at surfaces
by preparing the system at high temperatures with a small initial surface
magnetization. As in the bulk case, a power-law behaviour of the surface
magnetization is observed at early times, governed by the nonequilibrium
exponent $\theta_1=( x_i - x_1)/z$. Recently the two-time surface autocorrelation
function $C_1(t,s)$ was analysed in the short-time regime $t-s \ll s$.\cite{Ple03}
Using scaling arguments it was predicted that in the case $x_i < x_1$ a new effect, called
cluster dissolution, takes place, which should lead to an unconventional, 
stretched exponential
dependence of the short-time autocorrelation:
\begin{equation} \label{eq_strexp}
C_1(t-s) \sim  \exp(-\overline{C} (t-s)^\kappa)
\end{equation}
with 
\begin{equation} \label{eq_kappa}
\kappa=\frac{(x_1-x_i)d}{z(d-1)}.
\end{equation}
It follows that for $t-s \ll s$ a stationary behaviour should be
observed, as the autocorrelation in this case only depends on the
time difference $t-s$.
These predictions were verified by
simulating the dynamical evolution of the Hilhorst-van Leeuwen model for
various values of $x_1$. Interestingly, this stretched exponential behaviour is
also observed in the three-dimensional semi-infinite
Ising model at the ordinary transition \cite{Ple03} where $x_i = 0.74$ and $x_1 = 1.26$.
For the case $x_i > x_1$, on the other hand, no stationary behaviour is predicted
for $t-s \ll s$.
It is worth noting that dynamical surface response functions have not yet
been analyzed in critical semi-infinite systems.

In the dynamical scaling regime ($t,s,t-s \gg t_{micro}$) it is expected
that the surface autocorrelation function $C_1(t,s)$ and the surface autoresponse
function $R_1(t,s)$ exhibit a behaviour similar to that of the corresponding
bulk quantities, but with surface exponents replacing the bulk exponents:
\begin{eqnarray}
C_1(t,s) & = & s^{-b_1} \, f_{C_1}(t/s) \label{C1} \\
R_1(t,s) & = & s^{-1-a_1} \, f_{R_1}(t/s) \label{R1}
\end{eqnarray}
The scaling functions $f_{C_1}(y)$ and $f_{R_1}(y)$ display again a power-law behaviour
in the limit $y \longrightarrow \infty$:
\begin{equation} \label{fc1}
f_{C_1}(y) \sim y^{-\lambda_C^1/z} \;\; , \;\;
f_{R_1}(y) \sim y^{-\lambda_R^1/z}.
\end{equation}
In the case of a quench to a critical point from an initially uncorrelated state, one expects
\begin{equation} \label{lc1}
\lambda_C^1 = \lambda_R^1 = \lambda_1
\end{equation}
where $\lambda_1$ is given by Eq.\ (\ref{eq_lambda1}). Furthermore, one should have
that\cite{Die83} 
\begin{equation} \label{a1}
a_1=b_1 = 2 x_1/z 
\end{equation}
where $x_1$ is again the surface scaling dimension.
  
\section{Surface autocorrelation in the aging regime}
The scaling behaviour (\ref{C1}) of the surface autocorrelation function
is predicted to be valid in the dynamical scaling regime for all values of $x_1$. 
In the short-time regime, however, one
has to distinguish whether $x_i > x_1$, so that the
usual domain growth mechanism prevails (this is the same situation as encountered in
bulk systems), or whether $x_i < x_1$, in which case cluster dissolution
takes place,\cite{Ple03} leading to the stationary stretched exponential behaviour
(\ref{eq_strexp}). In the domain growth regime, the scaling form (\ref{C1})
is expected to be valid even for $t-s < s$.
 
These predictions are confronted in the following with results obtained
in extensive Monte Carlo simulations of the Hilhorst-van Leeuwen model and
of the three-dimensional semi-infinite Ising model. For the former model, various values
of the amplitude $A$ were considered, with the value of the scaling dimension
of the surface magnetization ranging from $x_1 = 1/4$ to $x_1 = 1$, see Eqs.\
(\ref{eq_hvl}) and (\ref{eq_hvl2}). The homogeneous two-dimensional
semi-infinite Ising model is recovered in the special case $A = 0$ and 
$x_1 = 1/2$. The different systems were prepared at $t=0$ at infinite
temperature (i.e. completely uncorrelated initial state) and then quenched
down to the critical point. Applying heat-bath dynamics, the dynamical evolution
of the system was studied for times $t > 0$. Heat-bath dynamics was
chosen so that no macroscopic quantities were conserved. After the waiting time
$s$ the time dependence of the dynamical surface autocorrelation function $C_1(t,s)$
was measured, with
\begin{equation} \label{c_ising}
C_1(t,s) = \frac{1}{N} \, \sum\limits_{i \in surface} \langle \sigma_i(t) \, \sigma_i(s) \rangle 
\end{equation}
and $t > s$.
The sum in (\ref{c_ising})
is over all surface spins, $N$ is the total number of surface sites,
whereas $\sigma_i(t)$ is the value of the spin located
at the surface site $i$ at time $t$. The brackets indicate an average over
thermal noise. Typically, the average has been taken
over more than 10000 different realizations
with different random numbers. In the actual simulations, systems with $300 \times
300$ spins were simulated in two dimensions, with periodic boundary condition
in the direction parallel to the surfaces and free boundary condition
in the non-homogeneous direction. For the three-dimensional Ising model,
systems with typically 40 layers with $60 \times 60$ spins per layer were studied.
Here, periodic boundary conditions were used in both directions parallel to the
surfaces. In both cases some simulations for other system sizes were also done, 
in order to check against finite-size effects. All the data discussed in this
work are free from this kind of effects.

\begin{figure}
\centerline{\epsfxsize=3.25in\ \epsfbox{
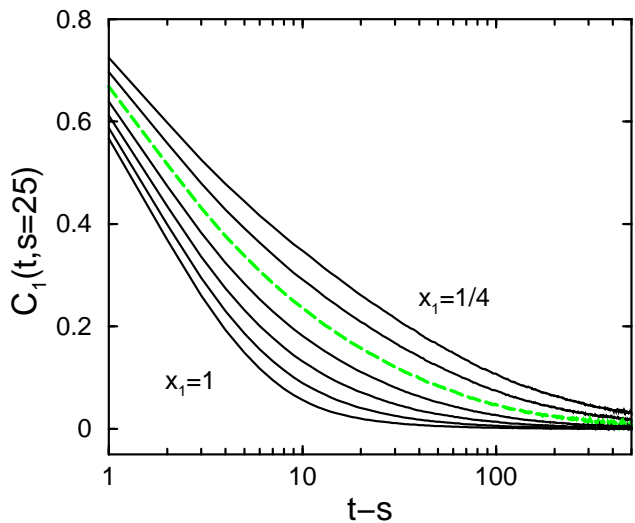}
}
\caption{Dynamical surface spin-spin autocorrelation functions vs
$t-s$ for $s$ fixed at 25, as obtained in the Hilhorst-van Leeuwen model. 
The different curves correspond 
to different values of the surface scaling dimension: $x_1 = 1/4$, 3/8, 1/2, 5/8, 3/4,
7/8, 1 (from top to bottom). The dashed grey line is obtained for the
pure homogeneous two-dimensional Ising model with $x_1 = 1/2$. 
}
\label{fig_1}
\end{figure}

Figure \ref{fig_1} gives a first impression of the impact that the surface scaling
dimension $x_1$ has on the surface two-time autocorrelation function. The Figure
displays autocorrelation functions obtained in the Hilhorst-van Leeuwen model
for various values of $x_1$, ranging from $x_1 = 1/4$ (top) to $x_1=1$ (bottom),
with the waiting time fixed at $s=25$. The grey dashed line is the function obtained
for $A=0$, i.e.\ for the homogeneous two-dimensional semi-infinite Ising model.
The data show that when the value of $x_1$ increases, the decay of the correlation
becomes faster. This is due to two different effects: (a) in the long-time limit,
the scaling function $f_{C_1}(t/s)$, which is proportional to $C_1(t,s)$, displays 
a power-law decay with the
exponent $\lambda_1/z$, where $\lambda_1$ increases with $x_1$ (see Eq.\
(\ref{eq_lambda1})), and (b) at short times with $x_1 > x_i$, the initial behaviour
is that of a stretched exponential where the exponent $\kappa$ again increases
with $x_1$,\cite{Ple03} see Eq.\ (\ref{eq_kappa}).

\begin{figure}
\centerline{\epsfxsize=3.25in\ \epsfbox{
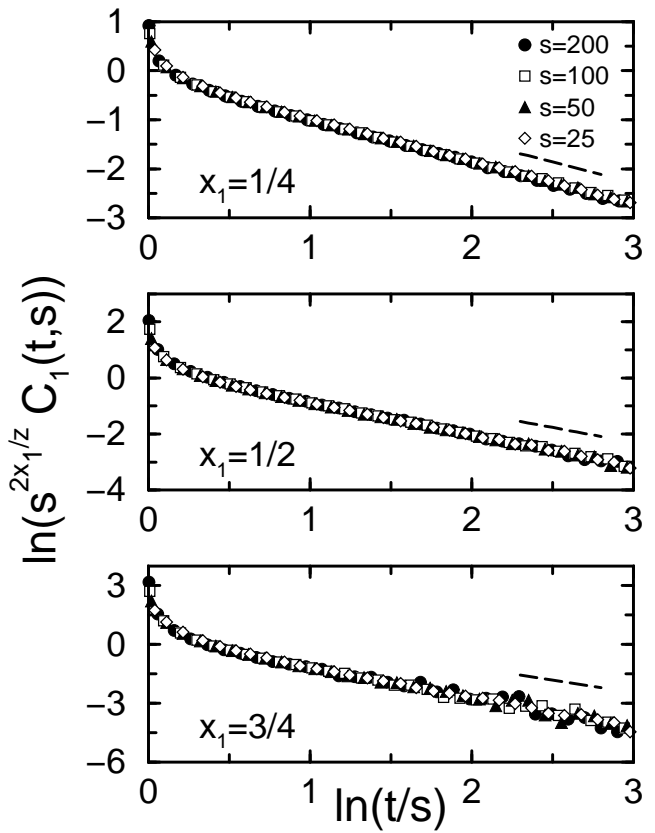}
}
\caption{Logarithm of the scaling function $f_{C_1}(t/s) = s^{2x_1/z} \, C_1(t,s)$
as function of the logarithm of the scaling variable $t/s$.
The data shown have been obtained in the Hilhorst-van Leeuwen model for three different
values of the surface scaling dimension $x_1$ and various waiting times $s$. The dashed lines
indicate the expected power-law behaviour for $t/s \gg 1$.
Here and in the following, errors are comparable to the scattering of the data.
}
\label{fig_2}
\end{figure}

\begin{table}
\caption{Comparison of numerically determined values of $\lambda_1/z$
with the theoretical expectations (\ref{eq_lambda1}). $\lambda_1/z$ has been
determined from both the autocorrelation ($C$) and the autoresponse ($R$)
functions. The errors result from averaging over different waiting times.
The available numerical data do not permit to reliably determine $\lambda_1/z$
for $x_1 > 3/4$ in the case of the Hilhorst-van Leeuwen model.
HvL: Hilhorst-van Leeuwen
model, OT: ordinary transition, SP: special transition point.
\label{table_2}}
\begin{tabular}{|c|c|c|c|c|c|}  \hline
HvL & $A$ &  $x_1$ & expected & $\lambda_1/z(C)$ & $\lambda_1/z(R)$ \\ \hline
& $0.50$ & $1/4$ & 0.85 & 0.84(1) & 0.84(1) \\
& $0.25$ & $3/8$ & 0.96 & 0.96(1) & 0.95(1)\\
& $0 $ & $1/2$ & 1.08 & 1.09(1) & 1.12(2)\\
& $-0.25$ & $5/8$ & 1.19 & 1.22(2) & 1.21(2) \\
& $-0.50$ & $3/4$ & 1.31 & 1.30(2) & 1.35(3)\\ \hline
$d=3$ & OT & $1.26$ & 2.10 & 2.10(1)& 2.18(3) \\
$d=3$ & SP & $0.376$ & 1.22 & 1.16(2) & 1.24(2)\\ \hline
\end{tabular}
\end{table}

The scaling function $f_{C_1}(t/s) = s^{2x_1/z} \, C_1(t,s)$ is shown in Figure
\ref{fig_2} for three different values of the scaling dimension $x_1$. In all cases
a perfect data collapse is achieved when plotting $f_{C_1}(t/s)$ for different waiting times.
As $x_1 > x_i$ for $x_1 =3/4$, one has in that case a stationary stretched exponential behaviour
at early times,\cite{Ple03} and therefore the scaling is not expected to work
at short times.
This is indeed the case, but this effect is not visible on the scale used in 
Figure \ref{fig_2} (see, however, the inset in Figure \ref{fig_4} below). 
The range of waiting times shown here is comparable to that 
of Ref.\ \onlinecite{Hen01}, where aging processes taking place in critical bulk systems
were investigated. The dashed lines illustrate the expected long-time behaviour
(\ref{fc1}) of the scaling function, which is in complete agreement with 
the numerical data, see Table \ref{table_2}. 
Unfortunately, no theoretical prediction of the complete 
functional form of the scaling function exists. There has been some recent 
progress in the derivation of an analytical expression for $f_{C}$ in the 
bulk system,\cite{Pic04,Hen04} but these
achievements are up to now only applicable to the special case $z=2$ not realized
in the critical systems we are interested in. 

\begin{figure}
\centerline{\epsfxsize=3.25in\ \epsfbox{
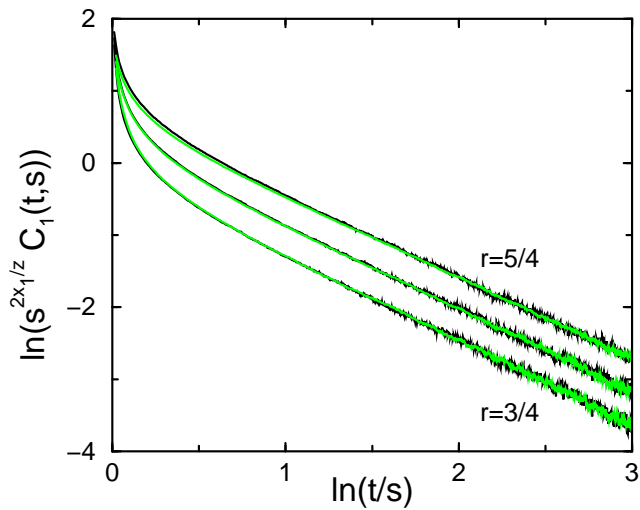}
}
\caption{Logarithm of the scaling function $f_{C_1}(t/s) = s^{2x_1/z} \, C_1(t,s)$ 
obtained in the
two-dimensional semi-infinite Ising model as function of $\ln (t/s)$ for three
different values of the strength $J_s = r \, J_b$ of the surface couplings,
with $r=5/4$ (top), 1, and 3/4. The black (grey) lines correspond to waiting times
$s=100$ ($50$).
}
\label{fig_3}
\end{figure}

In the three-dimensional semi-infinite Ising model different surface universality
classes are encountered when changing the strength $J_s$ of the surface couplings. This is
not the case in two dimensions, where one has only the ordinary transition
for all values of $J_s$. Nevertheless, the dynamical two-time autocorrelation
function is affected by a change of the value of $J_s$. This is illustrated in
Figure \ref{fig_3} where the logarithm of the scaling function
is plotted for three different
values of $r=J_s/J_b$: 5/4, 1, 3/4, and two different waiting times. 
One remarks that $C_1(t,s)$
decreases with increasing value of $r$ for a given value of $t/s$.
However, when multiplying $C_1(t,s)$ by
a constant so that the different curves coincide for large values of $t/s$, one
observes that the rescaled functions are identical
in the whole aging regime. It is only at short times with $t-s < s$ that
slight deviations are observed. This is not surprising, as the short-time
behaviour of the dynamical autocorrelation function is not expected to be 
universal in the present case. Concerning the aging regime, one may conclude
that the scaling law (\ref{C1}) should in fact read
\begin{equation} \label{eq_csurfscal2}
C_1(t,s) = \Omega(J_s) \,  s^{-2x_1/z} \, f_{C_1}(t/s)
\end{equation}
where only the amplitude $\Omega$ depends on the strength $J_s$ of the surface couplings.
The same conclusion holds for values of $A \neq 0$.

\begin{figure}
\centerline{\epsfxsize=3.25in\ \epsfbox{
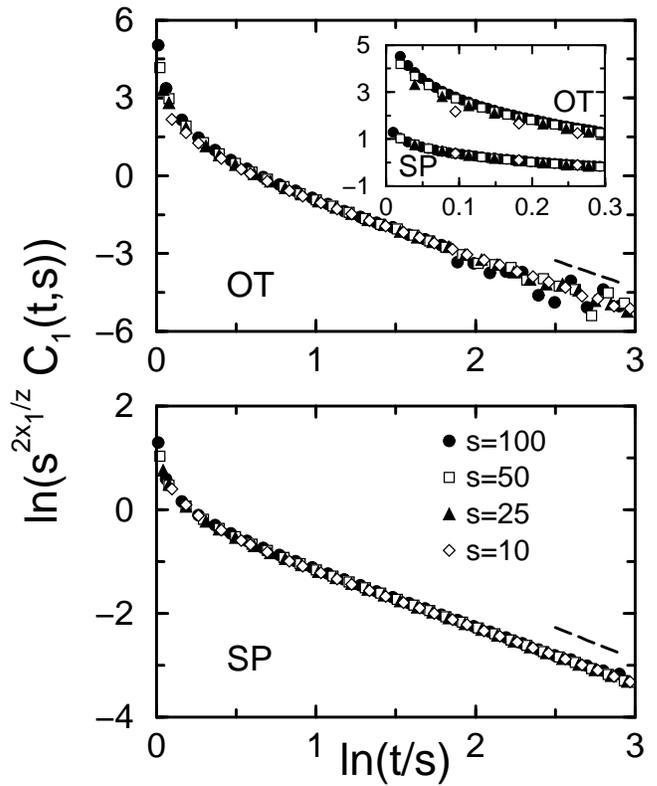}
}
\caption{The same as Figure \ref{fig_2}, but now for the three-dimensional Ising model
at the ordinary transition (OT) and at the special transition point (SP).
The dashed lines indicate again the expected power-law behaviour for $t/s \gg 1$.
Inset: short-time behaviour showing the predicted deviations from scaling
at the ordinary transition, whereas at the special transition point the scaling
behaviour (\ref{C1}) is also observed at short times.
}
\label{fig_4}
\end{figure}

Finally, Figure \ref{fig_4} shows the scaling functions $f_{C_1}$ determined in 
the semi-infinite three-dimensional Ising model for two different surface universality classes:
the ordinary transition (with $J_s = J_b$) and the special transition point
(with $J_s = 1.5 J_b$).
In both cases data collapse is achieved when inserting the correct
values of the surface scaling dimension, see Table \ref{table_1}. 
The inset illustrates
the short-time behaviour. Recall that at the ordinary transition $x_1=1.26 > x_i$,
whereas at the special transition point $x_1= 0.376 < x_i$. Following the
scaling arguments given above, a stationary behaviour is expected in the former
case, yielding at short times deviations from the scaling behaviour
(\ref{C1}), whereas in the latter case the same scaling behaviour 
should also prevail at short times. This is indeed what is observed.
 
\section{Surface thermoremanent magnetization}
In order to compute the surface thermoremanent magnetization we follow Barrat \cite{Bar98}
and prepare the system in an uncorrelated initial state before quenching it down
to the critical point in presence of a small binary random field $h_i = \pm h$. After
the waiting time $s$ the field is switched off and the surface (staggered)
magnetization 
\begin{equation} \label{eq_M}
M_1(t,s)= \frac{1}{N} \, \sum\limits_{i \in surface} \overline{\langle
h_i \, \sigma_i(t) \rangle}/T_c 
\end{equation}
is measured,
with $t > s$ being the time elapsed since the quench.
Here $\langle \cdots \rangle$ indicates again an average over the thermal noise 
whereas the bar means an average over
the random field distribution. The sum in Eq.\ (\ref{eq_M}) is restricted to
the $N$ surface sites.

It is very expensive to obtain good data for the surface integrated response function
at larger values of the ratio $t/s$. This is due to the value of the surface
scaling dimension, which governs to a large extent the behaviour in the 
aging regime and which in general considerably exceeds the value of the
scaling dimension in the bulk. In the simulations I went up to observation times
$t = 21 \, s$ and typically averaged over 500000 different realizations for every
waiting time considered. The sizes of the systems are the same as those used
for the study of the surface autocorrelation function.

The local thermoremanent magnetization is related to the surface response function
by
\begin{equation} \label{eq_M_2}
M_1(t,s) = h \, \int\limits_0^s \!\D u\, R_1(t,u).
\end{equation}
$M_1(t,s)$ should have the following scaling behaviour in the dynamical scaling regime
\begin{equation} \label{eq_Mskal}
M_1(t,s)/h = s^{-2 x_1/z} f_{M_1}(t/s) 
\end{equation}
with the scaling function $f_{M_1}$. Expression (\ref{eq_Mskal}) results from
inserting the expected behaviour (\ref{R1}) of the surface response function
into Eq.\ (\ref{eq_M_2}).

Exact expressions for the scaling functions $f_{R_1}$ and $f_{M_1}$ are obtained by
assuming that local scale invariance \cite{Hen01,Hen02} holds.
As shown in the Appendix, one then obtains for the surface autoresponse the
expression
\begin{equation} \label{eq_fR}
f_{R_1}(y) = r_0 \, y^{1+2 x_1/z-\lambda_1/z} \, (y-1)^{-1-2 x_1/z}
\end{equation}
and, after integration,
\begin{eqnarray} \label{eq_fM}
f_{M_1}(y)& = & r_0 \, y^{-\lambda_1/z} \nonumber \\
& & \hspace*{-2.3cm }_2F_1(1+2x_1/z, \lambda_1/z - 2x_1/z;
\lambda_1/z - 2x_1/z +1; y^{-1})
\end{eqnarray}
where $_2F_1$ is a hypergeometric function.
Expressions (\ref{eq_fR}) and (\ref{eq_fM}) are similar to those obtained
for bulk systems,\cite{Hen02} but now the surface scaling dimension $x_1$ and the surface
autoresponse exponent $\lambda_1$ replace the corresponding bulk quantities.

The scaling form (\ref{eq_Mskal}) results by naively inserting 
Eq.\ (\ref{R1}) into Eq.\ (\ref{eq_M_2}) without taking into
account that the condition for the validity of (\ref{R1}) is
violated at the lower integration bound. It was shown in \cite{Hen03} that
in case the system undergoes
phase-ordering at temperatures below $T_c$ the leading 
correction-to-scaling term can usually not be neglected. However, 
at a critical point this
correction term, which is of the form $s^{-\lambda_1/z} \, g_{M_1}(t/s)$,\cite{Hen03} where
$g_{M_1}$ is a scaling function known in the framework of the theory
of local scale invariance, should not have a sizeable effect,   
as $\lambda_1/z$ is much larger than $2 x_1/z$. (In case of a bulk system,
the local exponents $\lambda_1$ and $x_1$ have of course to be replaced
by the corresponding bulk exponents.) This has been confirmed
in the study of aging phenomena in critical bulk systems,\cite{Hen01}
and it is also the case for the surface thermoremanent magnetization discussed in the
following.

\begin{figure}
\centerline{\epsfxsize=3.25in\ \epsfbox{
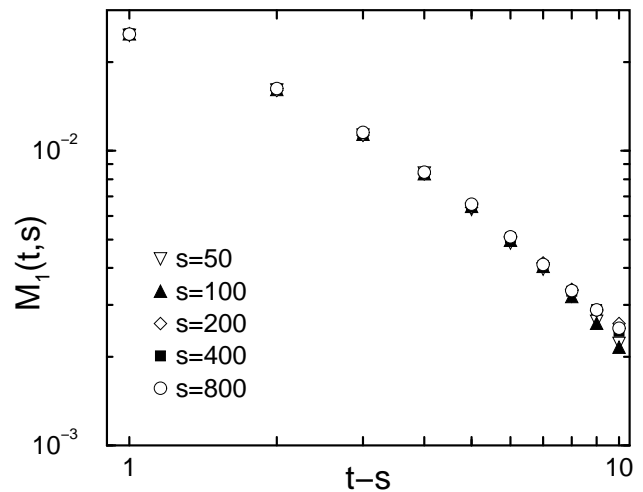}
}
\caption{
The thermoremanent magnetization as function of the 
time difference $t-s$ in the Hilhorst-van Leeuwen
model with $x_1=1$ and $h=0.1$. A stationary regime is observed at early times
as $M_1(t,s)$ only depends on $t-s$. In this figure error bars are comparable to the
symbol sizes.
}
\label{fig_5}
\end{figure}

The behaviour of the autoresponse function in the short-time regime
$t-s < s$ has not yet been discussed in
the literature.
Some conclusions regarding the autoresponse function may, however, be drawn by
remarking that for $t-s \ll s$ the system is in local equilibrium, yielding
a fluctuation-dissipation ratio $X(t,s)$, see (\ref{eq_x}), that takes on its equilibrium
value 1.\cite{Cri03} 
This ratio shows strong deviations from its equilibrium behaviour only at
later times when leaving the quasiequilibrium regime,
as discussed in the next Section.
It may therefore be concluded that the autoresponse function should exhibit at short times
a stationary behaviour every time the autocorrelation
function displays this kind of behaviour.
This is indeed the case, 
as illustrated in Figure \ref{fig_5}
for the Hilhorst-van Leeuwen model with $x_1=1$. 

\begin{figure}
\centerline{\epsfxsize=3.25in\ \epsfbox{
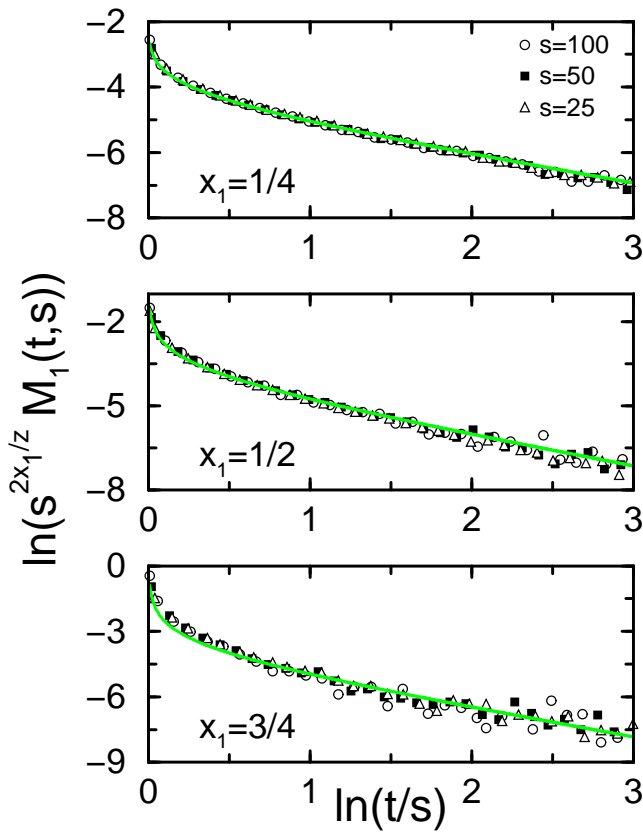}
}
\caption{
Logarithm of the scaling function of the thermoremanent magnetization
as function of $\ln(t/s)$, as obtained in the Hilhorst-van Leeuwen model for
three different values of the surface scaling dimension $x_1$ and for
different waiting times. The strength of
the binary random field was $h=0.1$. The full grey lines follow from the theoretical 
prediction (\ref{eq_fM}).
}
\label{fig_6}
\end{figure}

\begin{figure}
\centerline{\epsfxsize=3.25in\ \epsfbox{
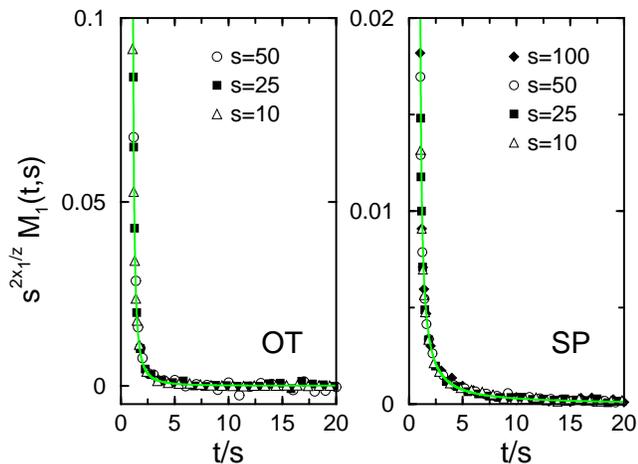}
}
\caption{
The scaling function of the thermoremanent magnetization vs $t/s$, as obtained
in the three-dimensional semi-infinite Ising model at the ordinary transition (OT) with
$J_s=J_b$ and at the special transition point (SP), with $h=0.05$. The full grey lines are
the theoretical predictions (\ref{eq_fM}) coming from the theory of local
scale invariance.
}
\label{fig_7}
\end{figure}

The main findings of this Section are summarized in Figures \ref{fig_6} and
\ref{fig_7}, showing the rescaled thermoremanent magnetization 
in the Hilhorst-van Leeuwen model for different
values of the surface scaling dimension as
well as in the three-dimensional Ising model at the ordinary transition
and at the special transition point. One remarks that in all cases the
thermoremanent magnetization shows a perfect scaling behaviour and that the
scaling function $f_{M_1}(t,s) = s^{2x_1/z} \, M_1(t,s)$ for $t-s > s$ 
indeed only depends on the 
scaling variable $t/s$. Deviations from scaling are observed 
for $t-s < s$ when $x_i < x_1$, as expected. 
The scattering of the data
in these figures being a measure of the error bars, it is obvious that reliable
data are more difficult to obtain for larger values of $x_1$.
This fact precludes the study of the autoresponse function in the aging regime
for values of $x_1$ much larger than 1.

The full lines in Figures \ref{fig_6} and \ref{fig_7} are obtained from the theoretical
prediction (\ref{eq_fM}) by inserting the numerical
values of the scaling dimensions $x_1$ and $x_i$. The non-universal amplitude $r_0$
has thereby be fixed by requiring that the analytical curve and the numerical data
coincide for large values of $t/s$. In cases where $x_i > x_1$ the theoretical
curves perfectly describe the numerical data over the whole range of values
of the ratio $t/s$. Deviations are observed in the short-time regime $t-s < s$
when $x_i < x_1$. This is obvious in Figure \ref{fig_6} for $x_1 = 3/4$,
but it is also the case for the three-dimensional Ising model at the ordinary
transition, as revealed by scrutinizing more closely the early time behaviour.
These deviations  
are of course due to the fact that for $t-s \ll s$, i.e. outside of the
dynamical scaling regime, the surface autoresponse (and therefore also
the thermoremanent magnetization) exhibits a stationary behaviour.
However, even in these cases the analytical
curve again nicely match the numerical data for $t-s > s$. 

The situation at the surface is therefore similar
to that encountered inside the bulk:\cite{Hen01}
in both cases no systematic deviations from the theoretical
prediction coming from the theory of local scale invariance are observed in
the aging regime when analyzing the numerical data. 
This is a remarkable result, as it indicates that
the scaling function of the autoresponse is completely fixed when knowing the
values of two critical exponents: the scaling dimension $x_l$ of the local
order parameter ($x_l$ being equal to $x$ resp.\ $x_1$ inside the bulk resp.\
close to the surface) and the scaling dimension $x_i$ of the initial
magnetization. At this stage one has to remember, however, that renormalization group
calculations \cite{Cal02} yield in the bulk system a correction to the
predicted behaviour (\ref{eq_lsi}) at two loops. Similar corrections are then
also expected in a field-theoretical treatment of the surface response function
in semi-infinite systems. It is an open problem why these corrections are
not directly revealed when comparing the numerical data with the local scale
invariance prediction. An obvious explanation would of course be that these
corrections are in fact very small and therefore difficult to observe.
There may, however, also be hidden a more fundamental problem. Indeed, exact results
in one dimension yield the value $\lambda_C=1$ for the Glauber-Ising model 
at the critical point $T=0$,\cite{Bra90}
whereas one obtains from the time-dependent Ginzburg-Landau equation,\cite{Bra95}
which is usually thought to describe the same system, the value $\lambda_C=0.6006$.
This indicates that the one-dimensional Glauber-Ising model and the
time-dependent Ginzburg-Landau equation belong to different universality
classes, which leaves open the possibility that similar problems 
arise in higher dimensions.\cite{Hen04b}
Clearly, further investigations are called for in order to clarify this
important issue.

Before coming to the surface fluctuation-dissipation ratio in the next Section,
let me briefly comment on the range of waiting times accessed in the present 
study. Indeed, a reader familiar with the investigations of aging phenomena
in spin glasses may be surprized by the apparently small values of $s$ retained
here. In critical ferromagnets the decay of, for example, the thermoremanent magnetization
is very fast, due to the large values of $\lambda_R/z$ (being
for the three-dimensional Ising model 1.73 in the bulk and 2.10 at the surface),
whereas in spin glasses the decay is much slower with $\lambda_R/z$ typically
of the order of $0.1 - 0.4$. This slow decrease makes it possible 
to access much longer
waiting times in spin glasses. 
On the other hand, however,
the dynamical scaling regime is entered for spin
glasses only for large values of $s$, whereas in ferromagnets dynamical
scaling behaviour is already observed for rather small values of $s$, as
illustrated for instance in the different figures discussed in this work. 

\section{Surface fluctuation-dissipation ratios}
One of the central quantity in the study of systems far from equilibrium
is the fluctuation-dissipation ratio $X(t,s)$ given by Eq.\ (\ref{eq_x}).\cite{Cri03} In the
context of critical systems one is especially interested in the limiting
value
\begin{equation}
X^\infty =
\lim\limits_{s \longrightarrow \infty} \left(
\lim\limits_{t \longrightarrow \infty} X(t,s) \right)
\end{equation} 
which characterizes the different universality classes.\cite{God00,Cal02} 
For example, one obtains in the case of Glauber dynamics
the value $X^\infty = 0.33$ for the two-dimensional Ising model,\cite{May03,Sas03}
whereas in 
three dimensions one has $X^\infty \approx 0.4$.\cite{God00}
In the exactly solvable one-dimensional case this value is 
$X^\infty = 1/2$ at $T=0$,\cite{God00a,Lip00} when starting from a uncorrelated
initial state.\cite{Hen03b}
This limiting value can also be obtained from the thermoremanent magnetization.\cite{God00}
Indeed, in the scaling limit the scaling functions $f_C$ and $f_R$ of the
autocorrelation and response functions only depend on $t/s$, yielding $M(t,s) =
M(C(t,s))$ and therefore\cite{God00}
\begin{equation} \label{x_inf}
X^\infty =
\lim\limits_{C \longrightarrow 0} \frac{T_c \, M(C)}{h \, C}.
\end{equation}
Fluctuation-dissipation ratios have up to now only been studied
in bulk systems. 

We define the surface fluctuation-dissipation ratio in complete analogy to the
bulk case, but with surface quantities replacing the bulk ones: 
\begin{equation} \label{eq_X1}
X_1(t,s)=T_c \, R_1(t,s)/\frac{\partial\, C_1(t,s)}{\partial \, s} 
\end{equation}
The limiting value of the surface fluctuation-dissipation ratio is then given
by the expression 
\begin{equation} \label{x1_inf}
X_1^\infty =
\lim\limits_{C_1 \longrightarrow 0} \frac{T_c \, M_1(C_1)}{h \, C_1}.
\end{equation}
where $C_1$ and $M_1$ are the surface autocorrelation function
and the surface thermoremanent magnetization studied in the
preceding Sections.

\begin{figure}
\centerline{\epsfxsize=3.25in\ \epsfbox{
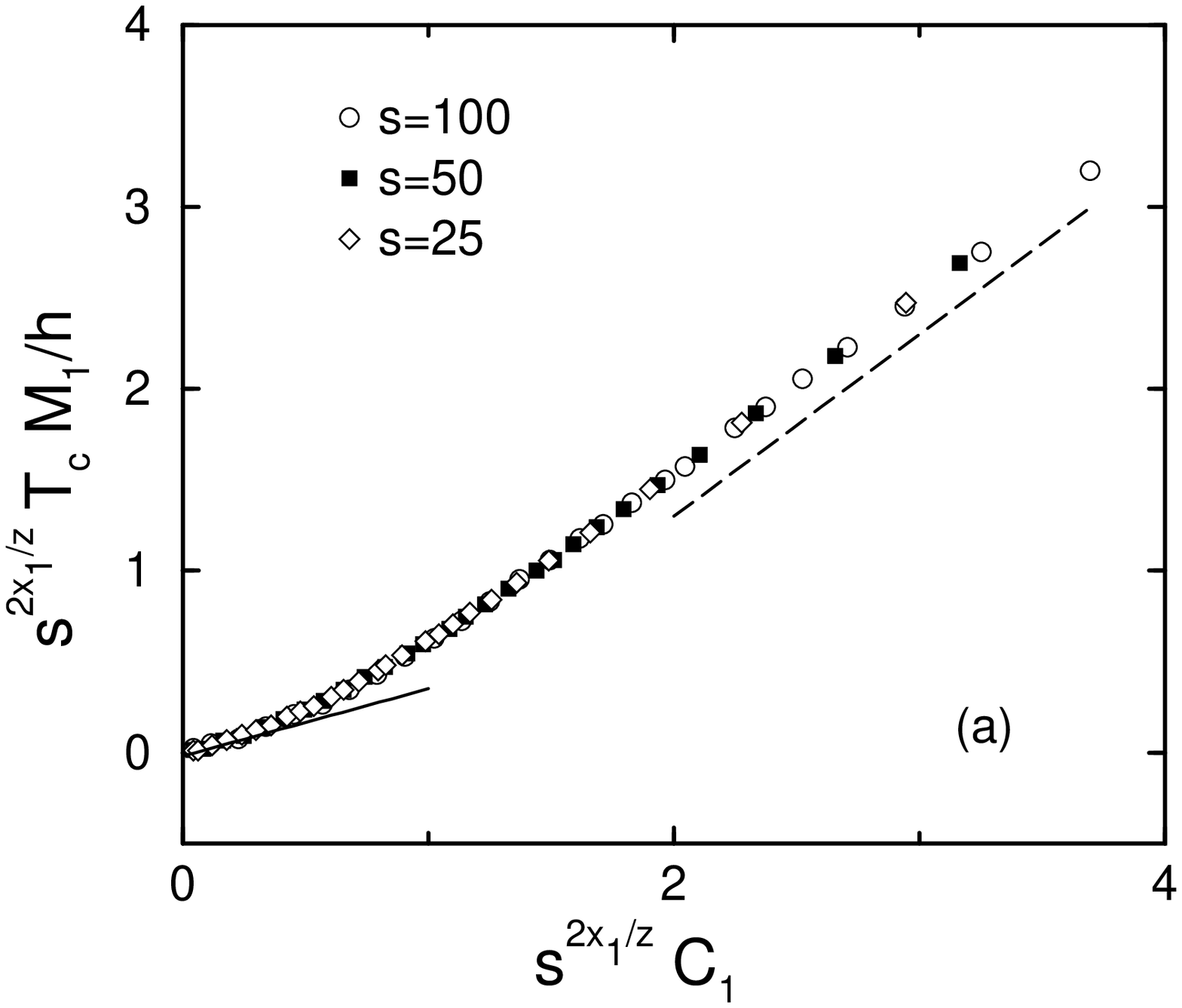}
}
\centerline{\epsfxsize=3.25in\ \epsfbox{
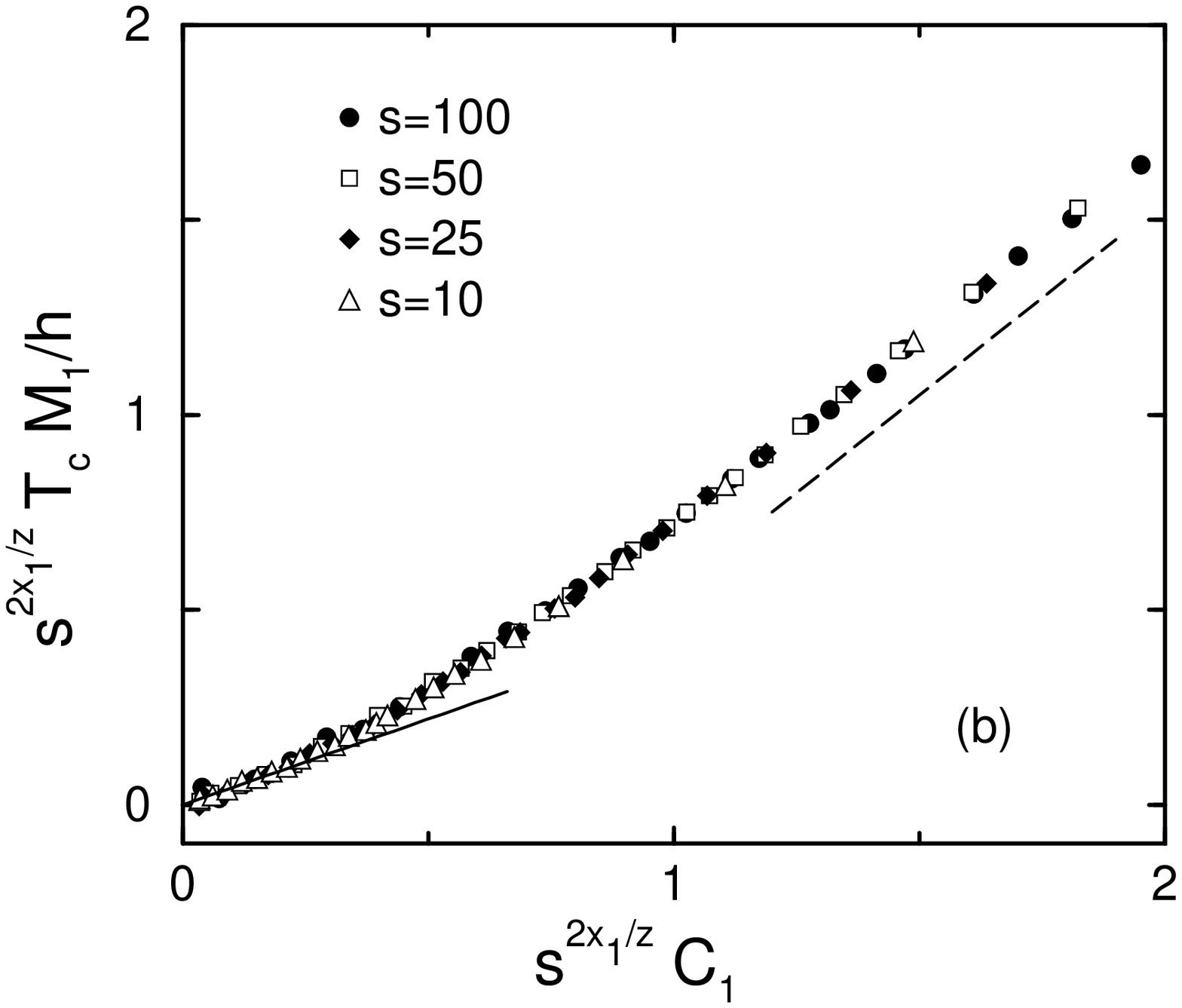}
}
\caption{
The rescaled surface thermoremanent magnetization versus the rescaled autocorrelation function:
(a) perfect two-dimensional semi-infinite Ising model with $A=0$ and (b) the three-dimensional
semi-infinite Ising model at the special transition point. The dashed lines indicates
the quasiequilibrium regime at early times, whereas the slope of the full lines is
given by the limiting value $X_1^\infty$ of the surface fluctuation-dissipation ratio,
see Eq.\ (\ref{x_inf}) and Table \ref{table_3}.
}
\label{fig_8}
\end{figure}

\begin{table}
\caption{The limiting value $X_1^\infty$ of the fluctuation-dissipation ratio
obtained from  Eq.\ (\ref{x_inf}) for the different models. HvL: Hilhorst-van Leeuwen
model, OT: ordinary transition, SP: special transition point. 
\label{table_3}}
\begin{tabular}{|c|c|c|c|}  \hline
HvL & $A$ & $x_1$ & $X_1^\infty$ \\ \hline
& $0.50$ & $1/4$ & $0.31(1)$ \\
& $0.25$ & $3/8$ & $0.33(1)$\\
& $0 $ & $1/2$ & $0.37(1)$ \\
& $-0.25$ & $5/8$ & $0.40(2)$ \\
& $-0.50$ & $3/4$ & $0.43(2)$ \\ \hline
$d=3$ & OT & $1.26$ & $0.59(2)$ \\
$d=3$ & SP & $0.376$ & $0.44(2)$ \\ \hline
\end{tabular}
\end{table}

Figures \ref{fig_8}a and \ref{fig_8}b display the parametric plot of the rescaled
thermoremanent magnetization $s^{2x_1/z} \, T_c\,  M_1/h$ versus the rescaled autocorrelation
function $s^{2x_1/z} \, C_1$ for two different cases: the pure two-dimensional semi-infinite
Ising model and the semi-infinite three-dimensional Ising model at the special transition
point. In both cases the data points for different waiting times show the expected
perfect scaling. A similar good scaling is observed for all studied systems in two
and three dimensions. At early times, $t-s \ll s$, a quasi-equilibrium behaviour,
where the fluctuation-dissipation theorem holds, is indeed observed, as indicated
by the dashed lines with slope 1. In the limit $C_1 \longrightarrow 0$ the slope
of these parametric plots yield the asymptotic values $X_1^\infty$ of the surface
fluctuation-dissipation ratios, as indicated by the full lines. 
Alternatively, one may also investigate $X_1$ as function of $s/t$, which yields the
limit value $X_1^\infty$ for $s/t \longrightarrow 0$. Both approaches give comparable results.
The limiting values
determined for the different cases are gathered in Table \ref{table_3}.
In the case of the Hilhorst-van Leeuwen model
the quality of the existing data do not allow a reliable determination
of $X_1^\infty$ for $x_1 > 3/4$.
The main message of Table \ref{table_3} is that $X_1^\infty$ depends on the surface scaling dimension:
for the Hilhorst-van Leeuwen model a variation in the strength of the defect
amplitude $A$ changes the value of $X_1^\infty$, whereas in the three-dimensional
Ising model the different universality classes yield a different value for this
limiting value. Assigning an effective local temperature via the equation $T_{eff} = T_c/X_1^\infty$,
one has to remark that (a) the surface effective temperature in general differs from
the effective temperature $T_c/X^\infty$ inside the bulk, and (b) the surface effective temperature  
depends on the surface quantities (as for example the strength of the surface couplings)
which determine the different surface universality classes. 

\section{Conclusion and outlook}
The aim of the present work has been to
extend the analysis of aging processes in critical systems
far from equilibrium to the semi-infinite geometry, paying special attention
to the behaviour of surface quantities in the dynamical scaling regime.
As shown in the preceding Sections, surface autocorrelation and autoresponse functions 
display in this aging regime a scaling behaviour similar to that observed in
the bulk, but with exponents differing from those encountered inside the bulk.
This scaling behaviour has been investigated numerically in the two- and
three-dimensional semi-infinite Ising models as well as in the Hilhorst-van Leeuwen
model. The latter model is a semi-infinite two-dimensional Ising model with an extended
surface defect which has the property that the surface scaling dimension
varies continuously as a function of the defect amplitude. It is this property
that has made possible a systematic study of the effects of surfaces on aging
processes.

It is remarkable that the scaling functions of the surface thermoremanent magnetization
are found to agree with the analytical predictions coming from the theory
of local scale invariance.\cite{Hen02} This extends the range of applicability
of this theory to semi-infinite systems. At this stage field-theoretical
computations of scaling functions are called for. Remembering the situation
in the bulk systems, one may expect that
renormalization group calculations also yield corrections to the predicted 
behaviour (\ref{eq_fR}) in semi-infinite systems. This (expected) discrepancy between the observed
agreement of the numerical data with the theoretical prediction on the one hand
and the appearance of correction terms in a field-theoretical treatment
on the other hand is an open problem which warrants more attention in the future.

Similar to what has been done in the bulk systems,\cite{Cal02} field theoretical
calculations should also permit to compute the limiting value $X_1^\infty$ of the surface
fluctuation-dissipation ratio. The dependence of $X_1^\infty$ on the surface scaling
dimension, as found in this work, 
again illustrates that the limiting value of the fluctuation-dissipation ratio permits
to characterize the different (surface) universality classes.  

Possible extensions of the present work include the study of nonequilibrium semi-infinite
systems in cases where the order parameter is conserved (Kawasaki dynamics).\cite{God04}
One would then be dealing
with the semi-infinite extension of model B (in the classification of Hohenberg, Halperin
and Ma), whereas in this work only the semi-infinite extension of model A has been considered.
Indeed, the study of semi-infinite critical systems evolving with Kawasaki dynamics gives rise to
new questions. It has been shown that in the semi-infinite extension of model B with
a conserved bulk order parameter the presence of nonconservative surface terms, leading
to a nonconserved local order parameter in the vicinity of the surface, yields
a different dynamic critical universality class as compared to the case
where these terms are absent. These two universality classes share the same
critical exponents but are characterised by different scaling functions
of equilibrium dynamic surface susceptibilities.\cite{Die94,Wic95} It is therefore
tempting to ask whether this different surface universality classes also show up
when studying nonequilibrium quantities. Especially, it would be interesting to
investigate the scaling functions of the autoresponse in these cases and to
compare them with the prediction (\ref{eq_fR}). A second obvious extension would be
the study of semi-infinite systems at temperatures below $T_c$ where phase-ordering
takes place. Dynamical scaling behaviour is also observed in bulk systems
in the low temperature phase and a similar behaviour is also expected
for semi-infinite systems. Work along the sketched lines is planed for the future.

\acknowledgments
I thank Malte Henkel and Ferenc Igl\'{o}i for inspiring discussions. The numerical
work has been done on the IBM supercomputer Jump at the NIC J\"{u}lich
(project Her10) and on the IA32 cluster of the Regionales Rechenzentrum Erlangen. I also
acknowledge support by the Bayerisch-Franz\"{o}sisches Hochschulzentrum
through a travel grant.

\appendix*
\section{}
In this Appendix I derive the expression (\ref{eq_fR}) for the scaling
function of the surface autoresponse under the assumption that local scale
invariance\cite{Hen02} holds. I hereby closely follow the derivation of
the scaling function of the bulk autoresponse given in Ref.\ \onlinecite{Hen02}.

Consider the local two-point function $\mathcal{R}_1(t,s;\vec{\rho}_1,\vec{\rho_2})$
which describes the response of the system at time $t$ at the surface site $\vec{\rho}_1$
to a perturbation acting at time $s$ at the surface site $\vec{\rho}_2$. As we
are considering aging systems, time translation invariance is broken. However we
still have space translation invariance along the surface, and therefore
$\mathcal{R}_1(t,s;\vec{\rho}_1,\vec{\rho_2})=\mathcal{R}_1(t,s;\vec{\rho})$ with
$\vec{\rho}= \vec{\rho}_1-\vec{\rho_2}$. We furthermore require that $\mathcal{R}_1$
transforms covariantly under scale ($X_0$) and special conformal ($X_1$) transformations,
i.e.\ $X_0 \, \mathcal{R}_1 = X_1 \, \mathcal{R}_1 = 0$, where the
differential operators $X_0$ and $X_1$ are explicitly given in Ref.\ \onlinecite{Hen02}.
The differential equations for the surface autoresponse $R_1(t,s) = 
\mathcal{R}_1(t,s;\vec{0})$ are obtained by setting
$\rho = | \vec{\rho} | = 0$, yielding
\begin{eqnarray}
& & \left( t \, \partial_t + s \, \partial_s + \zeta_1 + \zeta_2 \right) R_1(t,s)=0 \\
& & \left( t^2 \, \partial_t + s^2 \, \partial_s + 2 \zeta_1 t + 2 \zeta_2 s \right) R_1(t,s)=0
\end{eqnarray}
with the solution
\begin{equation} \label{anh_r1}
R_1(t,s)= r_0 \left( \frac{t}{s} \right)^{\zeta_2 - \zeta_1} (t-s)^{-(\zeta_1 + \zeta_2)} 
\, \Theta(t-s).
\end{equation}
Here, $\zeta_1$ and $\zeta_2$ are two exponents left undetermined by the theory, whereas $r_0$
is a normalization constant and $\Theta$ the step function. The exponents $\zeta_1$ and $\zeta_2$
are fixed
by comparing expression (\ref{anh_r1}) with the expected scaling behaviour
(\ref{R1}) and (\ref{fc1}), leading to the final result:
\begin{equation} \label{anh_r2}
R_1(t,s) =  r_0 \left( \frac{t}{s} \right)^{1+2x_1/z-\lambda_1/z} (t-s)^{-1-2x_1/z}\,  \Theta(t-s).
\end{equation}
This can be written in the form
\begin{equation}
R_1(t,s) =  s^{-1-2x_1/z} \, f_{R_1}(t/s)
\end{equation}
where the scaling function $f_{R_1}(y)$ is given by the expression (\ref{eq_fR}).




\begin{references}
\bibitem{God02} C. Godr\`eche and J.-M. Luck, J. Phys.: Condens. Matter
{\bf 14}, 1589 (2002).
\bibitem{Cug02} L.F. Cugliandolo, in {\it Slow Relaxation and
non equilibrium dynamics in condensed matter}, Les Houches Session 77 July 2002,
J-L Barrat, J Dalibard, J Kurchan, M V Feigel'man eds (Springer, 2003).
\bibitem{Cri03} A.\ Crisanti and F.\ Ritort, J.\ Phys.\ A {\bf 36},
R181 (2003).
\bibitem{Bra94} A.J. Bray, Adv.\ Phys.\ {\bf 43}, 357 (1994).
\bibitem{Cug94} L.F. Cugliandolo, J. Kurchan, and G. Parisi, J.\ Physique I {\bf 4},
1641 (1994).
\bibitem{Bar98} A. Barrat, Phys. Rev. {\bf E57}, 3629 (1998).
\bibitem{Ber99} L. Berthier, J.L. Barrat, and J. Kurchan, Eur. Phys. J.
{\bf B11}, 635 (1999).
\bibitem{God00a} C. Godr\`eche and J.M. Luck, J. Phys. A {\bf 33}, 1151 (2000).
\bibitem{God00} C. Godr\`eche and J.M. Luck,
J. Phys. A {\bf 33}, 9141 (2000).
\bibitem{Lip00} E. Lippiello and M. Zannetti, Phys.\ Rev.\ E {\bf 61}, 3369 (2000).
\bibitem{Hen01} M. Henkel, M. Pleimling, C. Godr\`eche, and J.-M. Luck,
Phys. Rev. Lett. {\bf 87}, 265701 (2001).
\bibitem{Ber01} L. Berthier, P.C.W. Holdsworth, and M.\ Sellitto,
J. Phys. A {\bf 34}, 1805 (2001). 
\bibitem{Cal02} P.\ Calabrese and A.\ Gambassi, Phys.\ Rev.\ B {\bf 65},
066120 (2002); Phys.\ Rev.\ E {\bf 66},
066101 (2002).
\bibitem{Hen02} M. Henkel, Nucl. Phys. {\bf B641}, 405 (2002).
\bibitem{Pic02} A. Picone and M. Henkel, J. Phys. A {\bf 35}, 5575 (2002).
\bibitem{May03} P. Mayer, L. Berthier, J.P. Garrahan, and P. Sollich,
Phys.\ Rev.\ E {\bf 68}, 016116 (2003).
\bibitem{Sas03} F. Sastre, I. Dornic, and H. Chat\'e, Phys.\ Rev.\ Lett.\
{\bf 91}, 267205 (2003).
\bibitem{Hen03c} M. Henkel and J. Unterberger, Nucl. Phys. {\bf B660},
407 (2003).
\bibitem{Hen03} M. Henkel, M. Pae{\ss}ens, and M. Pleimling,
Europhys. Lett. {\bf 62}, 664 (2003); Phys.\ Rev.\ E {\bf 69}, 056109 (2004).
\bibitem{Hen03a} M. Henkel and M. Pleimling, Phys.\ Rev.\ E {\bf 68}, 065101(R) (2003).
\bibitem{Maz04} G.F. Mazenko, Phys.\ Rev.\ E {\bf 69}, 016114 (2004).
\bibitem{Abr04} S. Abriet and D. Karevski, Eur. Phys. J. B 37, 47 (2004).
\bibitem{God04} C. Godr\`eche, F. Krz\c{a}kala, and F. Ricci-Tersenghi,
J. Stat. Mech.: Theor. Exp. P04007 (2004).
\bibitem{Pic04} A. Picone and M. Henkel, Nucl.\ Phys.\ B {\bf 688}, 217 (2004).
\bibitem{Cha04} C. Chatelain, cond-mat/0404017.
\bibitem{Abr04b} S. Abriet and D. Karevski, cond-mat/0405598.
\bibitem{Hus89} D.A. Huse, Phys. Rev. {\bf B40}, 304 (1989).
\bibitem{Jan89} H.K. Janssen, B. Schaub, and B. Schmittmann, Z. Phys. B{\bf 73}, 
539 (1989).
\bibitem{Cug97} L.F. Cugliandolo, J. Kurchan, and L. Peliti, Phys.\ Rev.\ E {\bf 55},
3898 (1997).
\bibitem{Hen03b} M. Henkel and G. Sch\"{u}tz, J. Phys. A {\bf 37}, 591 (2004).
\bibitem{Cha03} C. Chatelain, J. Phys. A {\bf 36}, 10739 (2003).
\bibitem{Die97} H.W. Diehl, Int.\ J.\ Mod.\ Phys.\ B {\bf 11}, 3503 (1997).
\bibitem{Ple04a} M. Pleimling, J.\ Phys.\ A {\bf 37}, R79 (2004).
\bibitem{Hil81} H.J. Hilhorst and J.M. van Leeuwen, Phys. Rev. Lett. {\bf 47}, 
1188 (1981).
\bibitem{Igl93} F. Igl\'oi, I. Peschel, and L. Turban, Adv. Phys. {\bf 42}, 683
(1193), and references therein.
\bibitem{Die83} S.\ Dietrich and H.W. Diehl, Z.\ Phys.\ B {\bf 51}, 343 (1983).
\bibitem{Die94} H.W. Diehl, Phys.\ Rev.\ B {\bf 49}, 2846 (1994).
\bibitem{Die02} H.W. Diehl, M. Krech, and H.\ Karl, Phys.\ Rev.\ B {\bf 66},
024408 (2002).
\bibitem{Rit95} U.\ Ritschel and P.\ Czerner, Phys. Rev. Lett. {\bf 75}, 3882 (1995).
\bibitem{Maj96} S.N. Majumdar and A.M. Sengupta, Phys. Rev. Lett. {\bf 76}, 2394 (1996).
\bibitem{Kik85} M.\ Kikuchi and Y.\ Okabe, Phys. Rev. Lett. {\bf 55}, 1220 (1985).
\bibitem{Rie85} H.\ Riecke, S.\ Dietrich and H.\ Wagner, Phys. Rev. Lett. {\bf 55}, 3010 (1985).
\bibitem{Ple03} M.\ Pleimling and F.\ Igl\'{o}i, Phys. Rev. Lett. {\bf 92},
145701 (2004).
\bibitem{Bin84} K. Binder and D.P. Landau, Phys. Rev. Lett. {\bf 52}, 318 (1984).
\bibitem{Rug92} C. Ruge, A. Dunkelmann, and F. Wagner, Phys. Rev. Lett. {\bf 69}, 2465
(1992); C. Ruge, A. Dunkelmann, F. Wagner, and J. Wulf, J.\ Stat.\ Phys. {\bf 73},
293 (1993).
\bibitem{Blo83} H.W.J. Bl\"{o}te and H.J. Hilhorst, Phys.\ Rev.\ Lett.
{\bf 51}, 2015 (1983).
\bibitem{Hoh77} P.C. Hohenberg and B.I. Halperin, Rev.\ Mod.\ Phys. {\bf 49}, 435 (1977).
\bibitem{Die92} H.W. Diehl and H.K. Janssen, Phys.\ Rev.\ A {\bf 45}, 7145 (1992).
\bibitem{Hen04} M.\ Henkel, A.\ Picone, and M.\ Pleimling, cond-mat/0404464.
\bibitem{Bra90} A.J. Bray, J. Phys. A {\bf 23}, L67 (1990).
\bibitem{Bra95} A.J. Bray and B. Derrida, Phys.\ Rev.\ E {\bf 51}, R1633 (1995).
\bibitem{Hen04b} M. Henkel, Adv. in Solid State Phys.\ {\bf 44} (2004) (in press)
and cond-mat/0404016.
\bibitem{Wic95} F. Wichmann and H.W. Diehl, Z.\ Phys.\ B {\bf 97}, 251 (1995).

\end{references}
\end{document}